\documentclass[%
superscriptaddress,
preprintnumbers,
 amsmath,amssymb,
 aps,
pra,
]{revtex4-1}

\usepackage{graphicx}
\usepackage{dcolumn}
\usepackage{bm}
\usepackage{lineno}
\usepackage{units}

\begin{document}
\author{M.~Turner\textsuperscript{*}}
\affiliation{CERN, Geneva, Switzerland}
\email{marleneturner@lbl.gov, now at Lawrence Berkeley National Laboratory, Berkeley, USA}
\author{P.~Muggli}
\affiliation{Max Planck Institute for Physics, Munich, Germany}
\author{E.~Adli}
\affiliation{University of Oslo, Oslo, Norway}
\author{R.~Agnello}
\affiliation{Ecole Polytechnique Federale de Lausanne (EPFL), Swiss Plasma Center (SPC), Lausanne, Switzerland}
\author{M.~Aladi}
\affiliation{Wigner Research Center for Physics, Budapest, Hungary}
\author{Y.~Andrebe}
\affiliation{Ecole Polytechnique Federale de Lausanne (EPFL), Swiss Plasma Center (SPC), Lausanne, Switzerland}
\author{O.~Apsimon}
\affiliation{Cockcroft Institute, Daresbury, UK}
\affiliation{Lancaster University, Lancaster, UK}
\author{R.~Apsimon}
\affiliation{Cockcroft Institute, Daresbury, UK}
\affiliation{Lancaster University, Lancaster, UK}
\author{A.-M.~Bachmann}
\affiliation{CERN, Geneva, Switzerland}
\affiliation{Max Planck Institute for Physics, Munich, Germany}
\affiliation{Technical University Munich, Munich, Germany}
\author{M.A.~Baistrukov}
\affiliation{Budker Institute of Nuclear Physics SB RAS, Novosibirsk, Russia}
\affiliation{Novosibirsk State University, Novosibirsk, Russia}
% ...
\collaboration{The AWAKE Collaboration}
\noaffiliation

\title{Experimental Study of Wakefields Driven by a Self-Modulating Proton Bunch in Plasma}

\begin{abstract}
We study experimentally the longitudinal and transverse wakefields driven by a highly relativistic proton bunch during self-modulation in plasma. We show that the wakefields' growth and amplitude increase with increasing seed amplitude as well as with the proton bunch charge in the plasma. We study transverse wakefields using the maximum radius of the proton bunch distribution measured on a screen downstream from the plasma. We study longitudinal wakefields by externally injecting electrons and measuring their final energy. 
Measurements agree with trends predicted by theory and numerical simulations and validate our understanding of the development of self-modulation. Experiments were performed in the context of the Advanced Wakefield Experiment (AWAKE) \cite{ROYAL}.
\end{abstract}

\keywords{Proton driven plasma wakefield acceleration, Seeded Self-Modulation}
\maketitle
Plasma wakefields can accelerate charged particles with gradients larger than \unit[1]{GeV/m}. These gradients greatly exceed those in metallic structures (\unit[$<$100]{MeV/m}). Wakefields are excited when, e.g., a relativistic particle bunch interacts with plasma; their amplitude depends on the bunch and plasma parameters. 

Bunches carrying very large amounts of energy (\unit[$>$100]{GeV} per particle and \unit[$>$100]{kJ}) can excite wakefields with GeV/m gradients over hundreds of meters. Such wakefields could accelerate witness bunches to $\sim$ TeV energies~\cite{PROTONDRIVEN}. High-energy drivers are available, e.g., proton bunches from the CERN Large Hadron Collider (LHC). However, their rms length ($\sigma_z$) is 6 to \unit[12]{cm}, much longer than the plasma electron wavelength ($\lambda_{pe}$) at plasma densities ($n_{pe}$) needed to reach accelerating fields \unit[$>$1]{GV/m}: \unit[$n_{pe}>10^{14}$]{cm$^{-3}$} and $\lambda_{pe}<\unit[4]{mm}$.
By adjusting the plasma wavelength to match the bunch length ($\lambda_{pe}\simeq\sqrt{2}\pi\sigma_z$) the amplitude of the excited wakefields would be \unit[$<$10]{MV/m}~\cite{DAWNSON}.

The same bunch can drive $\sim$GV/m field amplitudes after self-modulation (SM) in plasma~\cite{PoP2-1326,PoP4-1154,SMI1,SMI2,SSM,SMIRESULTS,Gross}. To design experiments based on this acceleration scheme -- with potential applications for high-energy physics~\cite{calwellHEP} --  understanding the development of the
SM of a charged particle bunch along the plasma is important. The
SM process grows from seed wakefields and reaches saturation~\cite{SSMMarlene}. Ideally this process would be directly studied by changing the plasma length, as is for example done with free electron lasers by adjusting the effective undulator length~\cite{FEL}. 

Measurements presented in this paper were performed in the Advanced WAKefield Experiment (AWAKE)~\cite{AWAKEfac}. In AWAKE, the plasma length is fixed by the geometry of the vapor source and by the laser ionization process~\cite{SSM}. We therefore use variations of the input parameters -- in this case the seed timing along the bunch -- to show that the measured output parameters of the drive and accelerated bunches are in qualitative agreement with those predicted by numerical simulations. Our results indicate that SM saturation does occur before the end of the plasma column. The results presented in this paper are thus an important ingredient for the planning of future experiments, in particular the length of the self-modulator plasma for next AWAKE experiments~\cite{bib:run2}.

The evolution of SM can be deduced from proton defocusing caused by transverse wakefields ($W_{r}$)~\cite{SSMMarlene}. We measure the time-integrated, transverse distribution of the self-modulated proton bunch downstream from the plasma exit~\cite{TurnerIPAC2017}. We show below that the maximum radius ($r_{max}$) of this distribution is proportional to the integral of the transverse plasma wakefields' amplitude during growth.

To study wakefields after the SM process developed, we externally inject electrons, accelerate them and measure their energy downstream of the plasma. While SM develops, the wakefields' phase velocity evolves and acceleration dynamics are complex~\cite{Wdecay}. However, once the phase velocity stabilises, electrons gain energy consistently according to the integrated longitudinal field amplitude they experience from there on. Electron energy measurements thus yield information on the amplitude of the longitudinal wakefields. The measurements are performed at a fixed (injection) time delay with respect to the seed timing.

In AWAKE, laser ionization of rubidium vapor creates the plasma~\cite{PLASMA1}. The \unit[$\sim$120]{fs}, \unit[$<$450]{mJ} laser pulse singly ionizes the rubidium atom (vapor density \unit[$10^{14}-10^{15}$]{cm$^{-3}$}) and creates a plasma with a radial extent larger than \unit[1]{mm} over a distance of \unit[10]{m}, the length of the vapor column. The plasma electron density is equal to the rubidium vapor density~\cite{karl}. To seed the SM process~\cite{LOTOVSEEDING,SSM}, we overlap in time and in space the short laser pulse ($ \ll \lambda_{pe}/c$) with the proton bunch. A sharp onset of plasma (relativistic ionization front) travels with the proton bunch and drives the seed wakefield (with an amplitude much larger than the noise expected in the system~\cite{LOTOVSEEDING}).

We vary the relative timing $t_{seed}$ between the ionization front (seed timing) and the proton bunch. At \unit[$t_{seed}=0$]{ps} the laser pulse overlaps in time with the center of the proton bunch and the seed wakefield' amplitude is maximum. By varying $t_{seed}$, we change the amplitude of the transverse seed wakefields $W_{r,seed}$ at the plasma entrance (\unit[$z=0$]{m})~\cite{LINEARTHEORY}. This changes not only the proton bunch density at the seed timing, which determines $W_{r,seed}$, but it also changes the number of protons between the seed timing and the electrons. Varying these two parameters together is not optimum to study wakefields, as both $W_{r,seed}$ and $N_p$ change the wakefields' growth along the bunch and plasma. However, changing $t_{seed}$ preserves the fundamental wakefields' theory parameters ($n_{pe}$, $k_{pe}\sigma_r$, etc.). Changing these parameters would complicate the analysis and the conclusions.

Experiments were performed with the following parameters. The plasma density is \unit[$(2.036 \pm 0.007)\times10^{14}$]{electrons/cm$^3$} and constant over the \unit[10]{m}-long plasma~\cite{ERDEM}. The drive bunch is produced by the CERN Super Proton Synchrotron (SPS) and each proton has a momentum of \unit[($400\pm0.4$)]{GeV/c}~\cite{SPS}. The bunch has a population of \unit[$(3.1 \pm 0.2)\times10^{11}$]{protons}, a rms length of \unit[$\sigma_z = (10.0 \pm 0.3)$]{cm} (or \unit[(333$\pm$10)]{ps}) and a radial rms size of \unit[$\sigma_r = (0.18 \pm 0.03)$]{mm} at the plasma entrance.

\begin{figure}[htb!]
    \centering
    \includegraphics[width=\columnwidth]{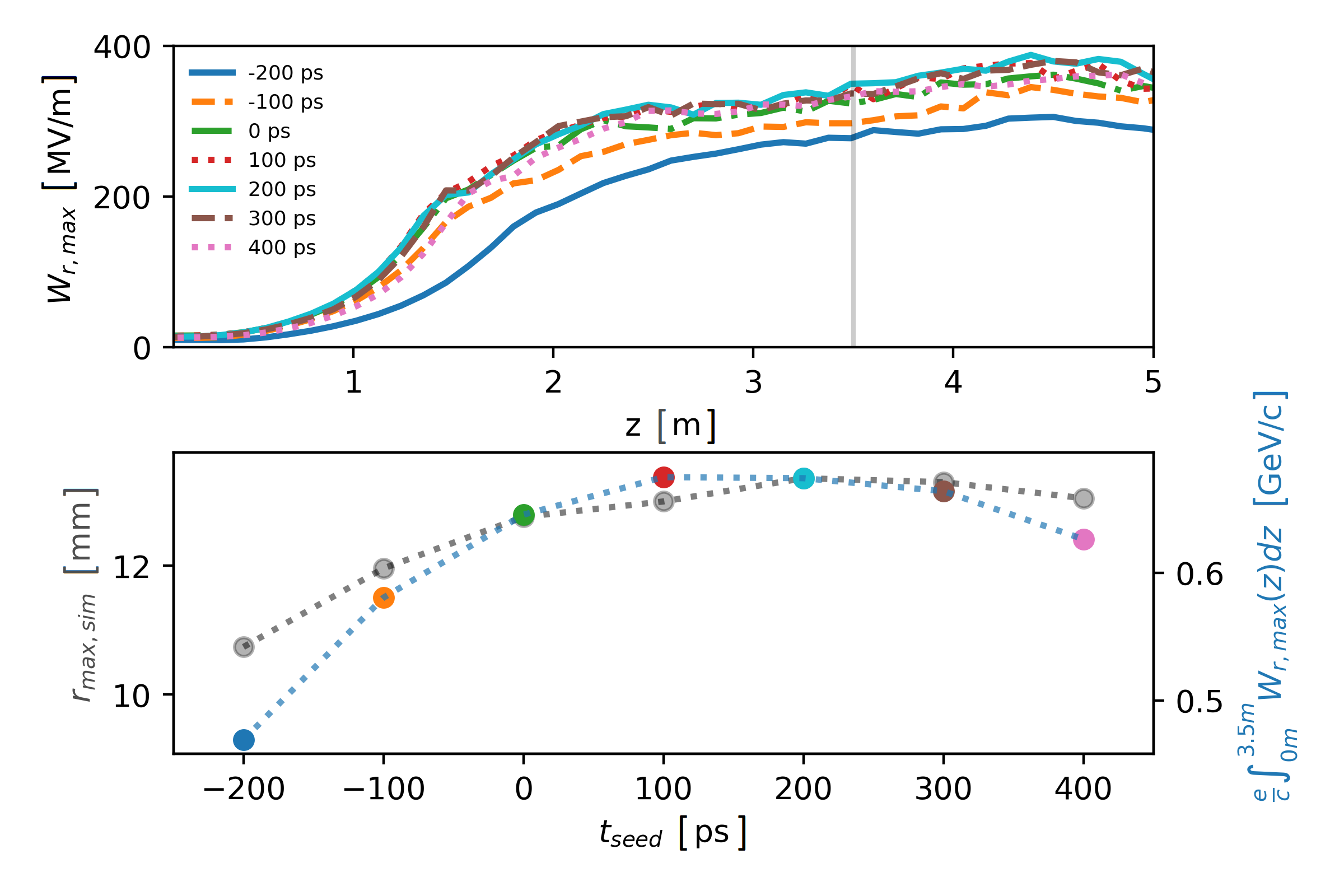}
    \caption{Top: maximum amplitude of the simulated transverse wakefields $W_{r}$ over the first \unit[5]{m} of plasma for different seed timings $t_{seed}$. The vertical line indicates the integration limit of \unit[3.5]{m}. Bottom: integral of the wakefields over the first \unit[3.5]{m} (symbols with same color as the corresponding line of the top graph, connected by the blue dotted line, right axis) and maximum radius of the simulated proton bunch transverse distribution $r_{max,sim}$ (after vacuum propagation to the measurement location, gray dots connected by the gray dotted line, left axis). The two curves are plotted with the same relative scaling (normalised to the maximum of each curve).}
    \label{fig:Wr}
\end{figure}

Protons are relativistic and the relative dephasing between them over \unit[10]{m} caused by energy gain or loss in the wakefields is negligible. During growth of the self-modulation, protons are focused and/or defocused. Defocused protons can radially exit the wakefields and
travel ballistically after that. Protons with the maximum transverse momentum ($p_{r,max}$) experienced the largest integral of transverse wakefields' amplitude and interaction distance: $p_{r,max}=\frac{e}{c}\int_0^{exit} W_r(r,z) dz$.

We performe 2D cylindrical, quasi-static simulations using LCODE~\cite{LCODE,LCODEN,LCODE2} with the experimental bunch and plasma parameters as input. The proton bunch is initialized with longitudinal and transverse Gaussian density profiles. We simulate seed timings $t_{seed}:$ $-200, -100, 0, 100, 200, 300$ and \unit[$400$]{ps}. For positive $t_{seed}$ values the seed is ahead of the middle of the proton bunch, for negative values behind it. In these simulations laser ionization is not simulated. Instead, the seeding is modelled by a step function cut of the proton bunch distribution at $t_{seed}$.

Simulation results show that for all these cases, protons that gain large transverse momentum radially exit the plasma between $z=3$ and \unit[4]{m} from the entrance. Thus, for each $t_{seed}$ we integrate the transverse wakefields' amplitude over the first \unit[3.5]{m} of plasma and evaluate the time along the bunch $\xi_{max}$ (from $t_{seed}$) for which the integrated field strength is maximum, i.e., the time along the bunch at which protons can acquire the maximum transverse momentum. We obtain \unit[$\xi_{max}\cong 317, 317, 317, 317, 330, 380, 420$]{ps} for \unit[$t_{seed}:$ $-200, -100, 0, 100, 200, 300$ and $400$]{ps} \cite{EXP1}. Transverse wakefields are evaluated at their radial maximum. In reality, protons start from an initial Gaussian distribution and travel radially across the wakefields. This calculation therefore yields an upper limit for the acquired transverse momentum. The top plot of Fig.~\ref{fig:Wr} shows the maximum amplitude of the transverse wakefields $W_{r,max}$ along the first \unit[5]{m} of plasma. We then integrate these wakefields along the plasma ($\frac{e}{c}\int_0^{3.5m}W_{r,max}(z)dz$) and plot the resulting values with points of the same color on the bottom plot of Fig.~\ref{fig:Wr}. They represent the largest momentum protons could gain during the growth of the SM process. It is clear that momentum gain by individual protons depends on their actual location both along and across the bunch and wakefields. 

Figure~\ref{fig:Wr} shows that the fields, as well as their integrated values, are smallest for negative seed times (\unit[$t_{seed}=-200, -100$]{ps}).  These cases correspond to the smallest number of protons 
behind the seed point
and driving wakefields. They also have the smallest wakefields' growth along the plasma \cite{SMI2,GROWTH2}. While the \unit[$\pm200$]{ps} curves have a similar seed wakefields' amplitude $W_{r,seed}=W_{r,max}(\xi=\xi_{seed})$, $W_{r,max}$ reaches only \unit[$\sim$260]{MV/m} for \unit[$t_{seed}=-200$]{ps}, but \unit[$\sim$330]{MV/m} for \unit[$t_{seed}=+200$]{ps} (due to the difference in $N_p$). Figure~\ref{fig:Wr} shows that the integrated values also tend to decrease for more forward seed times (\unit[$>200$]{ps}). This is consistent with the fact that the amplitude of the initial seed field decreases when seeding ahead of the center of the bunch.

\begin{figure}[htb!]
    \centering
    \includegraphics[width=\columnwidth]{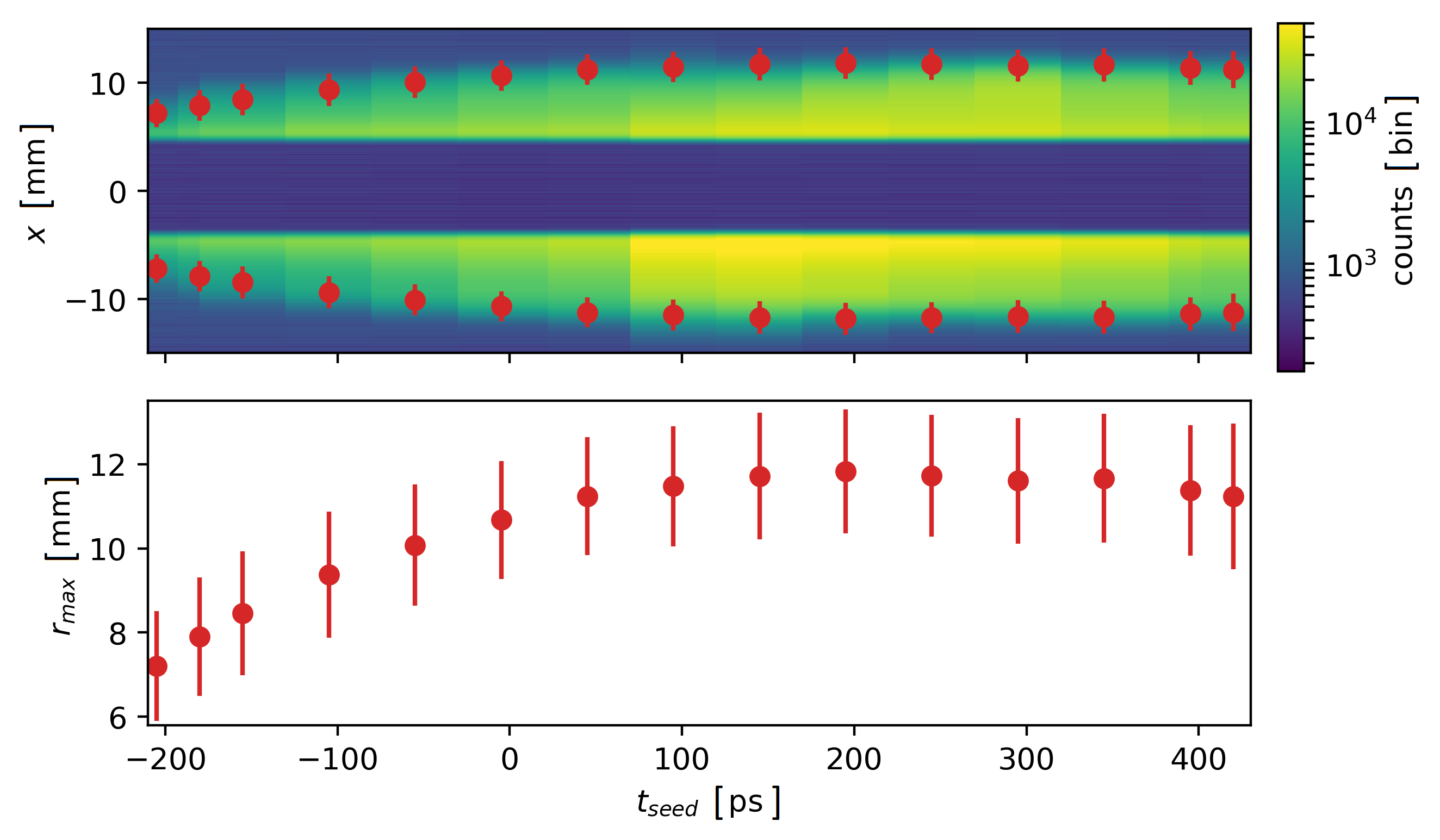}
    \caption{Top: summed waterfall plot of the horizontal line-outs of the measured self-modulated proton bunch transverse distribution as a function of $t_{seed}$. Note that the intense light emitted by focused protons (protons in the core, \unit[-3$\le x\leq+5$]{mm}) is blocked by a mask \cite{TurnerIPAC2017}. Each vertical line of pixels shows the average of all measurement at a given $t_{seed}$. Note the logarithmic color scale. Top and bottom: maximum radius $r_{max}$ (red dots) of the proton bunch distribution. The points show the average of the measurements and the error bars show the standard deviation of the individual measurements linearly combined with the resolution of the camera (\unit[0.1]{mm}) and the screen (\unit[0.3]{mm}).}
    \label{fig:maxr}
\end{figure}

In simulations, we propagate the self-modulated proton bunch in vacuum from the plasma exit to the location where we measure the transverse, time-integrated 
proton bunch distribution in the experiment, i.e., \unit[10]{m} downstream. We identify the maximum displacement of the protons $r_{max,sim}$ for each $t_{seed}$, and plot these values as gray dots on the bottom plot of Fig.~\ref{fig:Wr}, to be compared with Fig.~\ref{fig:maxr}. 

Both the maximum radius and the integral value of the transverse wakefields' amplitude depend on the wakefields' seed value and growth along the plasma. The similarity of the shape of the two curves demonstrates that in simulations the maximum radius is a measure of the integral of the transverse wakefields' amplitude. This shows that the details regarding how protons acquire transverse momentum and where they exactly exit the wakefields are not important. Integrating transverse wakefields over the exact distance protons experience wakefields slightly changes final values, but not the general shape of the curve. This enables us to compare simulations and experimental data since only the maximum radius of the proton bunch distribution is measured in the experiment and not the location where the protons radially exit the wakefields. 

In the experiment (Fig.~\ref{fig:maxr}), we changed the seed time $t_{seed}$ from \unit[$-205$]{ps} to \unit[+445]{ps} and measured the transverse, time-integrated distribution of the self-modulated proton bunch, using a scintillating screen and a CCD camera \cite{TurnerIPAC2017} located \unit[$\sim10$]{m} downstream from the plasma exit.  

The top of Fig.~\ref{fig:maxr} shows a summed waterfall plot of the measurements: all images at a given $t_{seed}$ are summed, integrated in the vertical direction and plotted as a function of $t_{seed}$. The light from the core of the bunch (\unit[$-3<x<+5$]{mm}) is blocked by a mask to better detect the lower level light corresponding to the defocused protons distribution (\unit[$x\leq-3$]{mm} and \unit[$x\geq+5$]{mm}).

We determine the maximum radius $r_{max}$ of the defocused proton distribution following the procedure described in \cite{TurnerEAAC2017} and plot it as a function of $t_{seed}$ (Fig.~\ref{fig:maxr}, bottom). Figure~\ref{fig:maxr} shows that $r_{max}$ increases from \unit[$(7.2\pm1.3)$]{mm} to \unit[$(11.8\pm1.5)$]{mm} when the seed time is varied from \unit[$-205$]{ps} to \unit[+195]{ps}. For seed times \unit[$>+200$]{ps}, $r_{max}$ decreases to \unit[$(11.1\pm1.6)$]{mm}. This experimental trend is consistent with the trend from simulation results presented on the bottom plot of Fig.~\ref{fig:Wr}. Increasing the number of protons in plasma $N_p$ increases the wakefields' growth and transverse amplitude (as seen on the top plot of Fig.~\ref{fig:Wr}); the growth is approximately constant for seed times \unit[$>$0]{ps} as the amplitude of the transverse seed wakefields $W_{r,seed}$ decreases, but $N_p$ increases.

The maximum radii obtained from the experiment follow the same trend as that obtained from numerical simulations, but are systematically smaller. This difference is likely due to the difference between actual experimental parameters (not always measured) and the parameters assumed for simulations (e.g., transverse extent of the plasma, proton bunch radius at plasma entrance, etc.). Additionally, it is much easier to identify the maximum radius from simulation data (outermost macro-particle) than from experimental data, due to various sources of background on experimental images (e.g., secondary particles) or to the camera minimum detection threshold. We can thus expect the radial position of the outermost proton 
to be larger than that of the distribution measured in the experiment. Note that these reasons may impact the absolute numbers or relative scaling, but not the overall trend. 

To study wakefields after the bunch has self-modulated, we externally inject electrons and measure their final energy~\cite{ILOVEEA}. The electrons are produced by a photo-injector \cite{ELECTRON}, have an energy of \unit[(18.6$\pm$0.1)]{MeV} and a bunch charge of \unit[$\approx 400$]{pC}. The electron bunch has a rms length of \unit[$\sigma_z \geq 5$]{ps}, on the order of the wakefields' period. We thus expect to capture electrons for each event, though with varying charge. For the results shown below, the captured charge is relatively low (\unit[$<$100]{pC}) and the bunch relatively long, these electrons can therefore be considered as test electrons, i.e., they do not alter the wakefields. The bunch has a radial extent of the order of a few hundred microns at the injection location. The injection angle of the electron beam is \unit[$\sim 1.5$]{mrad} with respect to the proton bunch propagation axis, an angle at which we observed consistent charge capture and acceleration \cite{AAC2018}. We set the electron beam trajectory to cross the proton beam trajectory near the plasma entrance (\unit[$z\sim0$]{m}).

The electron bunch has a constant delay of \unit[$\tau= (250\pm50$)]{ps} %
with respect to the ionizing laser pulse ($\tau= 0$, per definition, where $\tau$ is the relative time along the bunch). The uncertainty of \unit[$\Delta\tau = \pm50$]{ps} results from differences in path length of the electron bunch trajectory in the transport line upstream the plasma entrance, but is constant for all measurements.

The electron trajectory is fixed and the timing jitter between the laser pulse and electron bunch is smaller than \unit[10]{ps} from event to event. The proton bunch has an arrival rms time jitter with respect to the laser pulse of \unit[$\sim 15$]{ps}. This timing jitter is short when compared to the proton bunch duration and the growth time of the wakefields along the bunch. We thus expect it to have no significant effect on the energy gain and results reported here. %

\begin{figure}[htb!]
    \centering
    \includegraphics[width=\columnwidth]{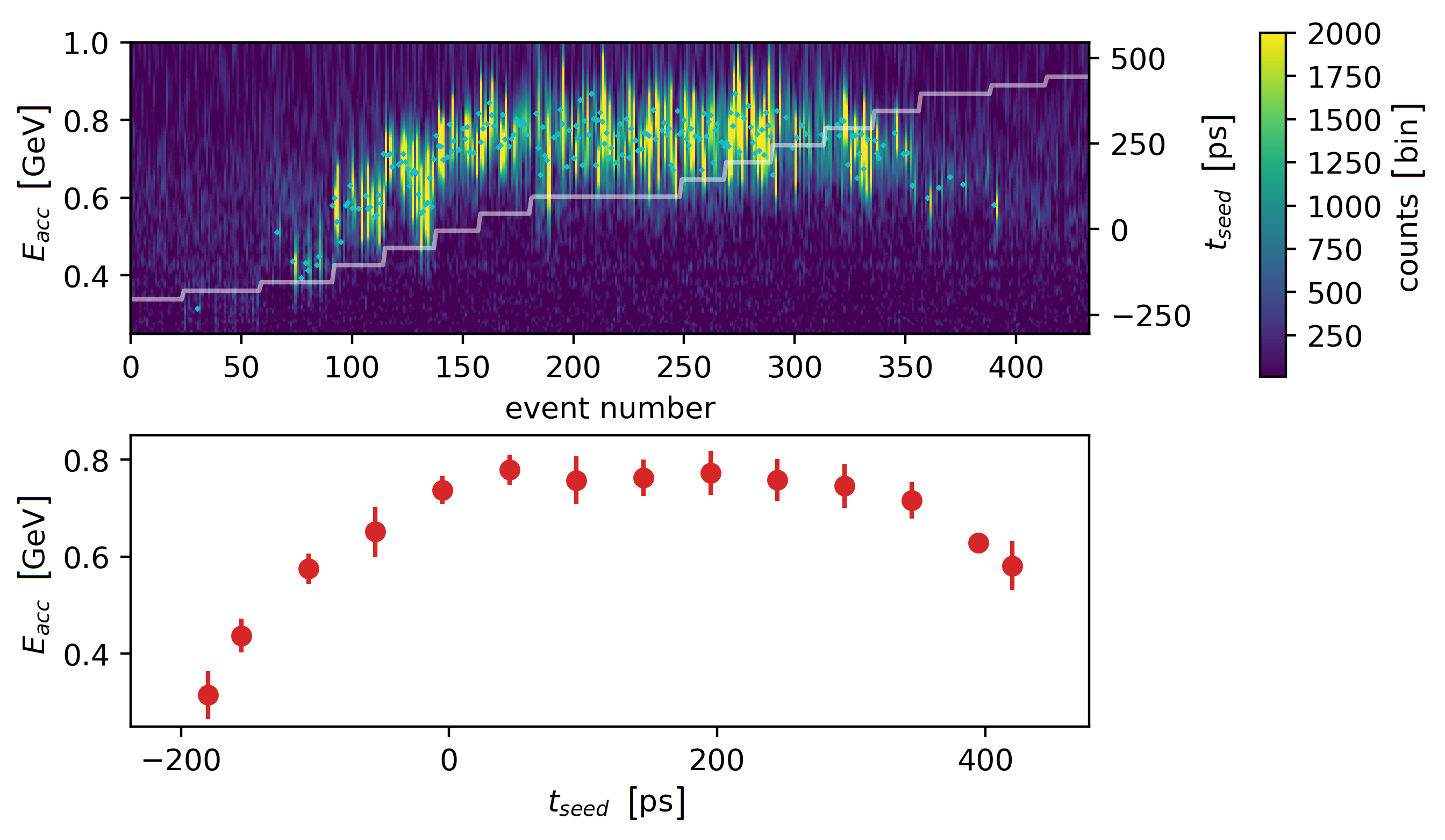}
    \caption{Top: waterfall plot of the measured electron energy spectra (right axis) during the seed scan. The value of $t_{seed}$ is shown by the white line and the vertical axis on the right. Cyan dots identify the charge peaks of the individual events. Bottom: energy of the charge peak of the accelerated electrons as a function of seed timing. The points show the average of the peak energy and the error bars show the standard deviation of individual measurements. %
    We note that the points at \unit[$t_{seed}\cong+390$]{ps} and \unit[$t_{seed}\cong -190$]{ps} are the result of a single measurement and have an error of \unit[$\approx$]{2\%}.}
    \label{fig:eacc}
\end{figure}

Electron spectra are acquired with an imaging magnetic spectrometer \cite{Spectro} for single events as two-dimensional images: energy in the dispersive plane, transverse size in the other. The energies quoted in the following are the values of the peak in the energy spectra obtained from images summed along the non-dispersive direction. Energy distributions have a finite width (much smaller than their average energy), visible on Fig.~\ref{fig:eacc} and similar to those in Ref.~\cite{ILOVEEA}. For the following measurements and simulations, we again varied $t_{seed}$ and kept all other parameters constant.

Figure~\ref{fig:eacc} (top) shows the vertical sum of the spectrometer images in a waterfall plot: each measurement corresponds to one vertical column (and energy spectrum) and is plotted as a function of event number. The value of $t_{seed}$ for each event is indicated by the white line. The measurement range was limited to \unit[$t_{seed}= -180$ to $+420$]{ps} by the accelerated electron charge that was not detectable beyond these values. Red dots and the corresponding error bars (bottom plot of Fig.~\ref{fig:eacc}) show the average energy and the standard deviation for all measurements at a given $t_{seed}$ value.
We see that the energy of the accelerated electrons reaches a maximum of \unit[$(0.77\pm0.05)$]{GeV} at seed times between 0 and \unit[$\sim$+200]{ps}. For seed times greater than \unit[$+200$]{ps} or smaller than \unit[$0$]{ps} the electron energy decreases to \unit[$\sim$0.6]{GeV} and \unit[$\sim$0.3]{GeV}, respectively.

The longitudinal wakefield' amplitude at the electron injection location depends on the seed wakefields' amplitude and on the number of protons between the seed and the electron bunch. Figure~\ref{fig:eacc} shows that the measured electron energy (and thus the average amplitude of the longitudinal wakefields experienced by electrons) decreases when seeding behind the center of the proton bunch.
The same is true for seed times earlier than \unit[$\sim$200]{ps}. We expect wakefields' amplitudes to be asymmetric around \unit[$t_{seed}=0$]{ps}, as: 1) moving the seed ahead of the proton bunch center first increases $N_{p}$ up to $t_{seed}=\tau/2$, and decreases it afterwards; moving the seed point backwards only decreases $N_{p}$; 2) the shape of the initial transverse seed wakefields includes the adiabatic response of the proton bunch and is thus different when the bunch envelope is decreasing or increasing along the bunch from the seed point. 

\begin{figure}[htb!]
    \centering
    \includegraphics[width=\columnwidth]{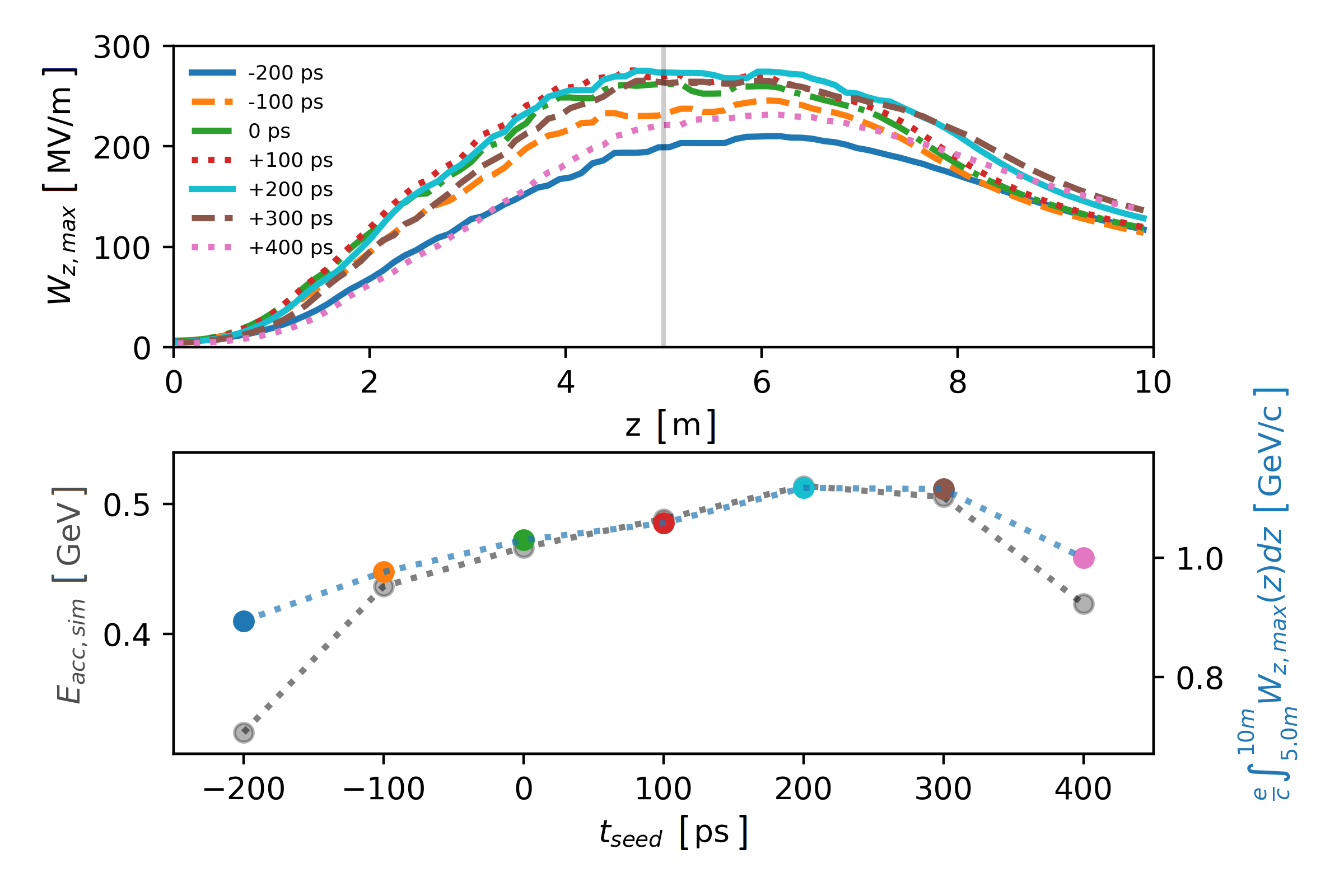}
    \caption{Top: amplitude of the simulated on-axis longitudinal wakefields along the plasma $W_{z}$ at \unit[$\tau=$250]{ps} for various $t_{seed}$. The vertical line indicates the lower bound of the integral. Bottom: integral from 5 to \unit[10]{m} of the fields from the top figure for the different $t_{seed}$ (symbols have the same color as the corresponding line and are connected by the blue dotted line, right $y$-axis). The gray points connected by the gray dotted line shows the observed electron energy in simulations (left $y$-axis). The two curves are plotted with the same relative scaling (normalised to the maximum of each curve).}
    \label{fig:eacc_sim}
\end{figure}

Again, we perform numerical simulations to obtain longitudinal wakefields' amplitudes ($W_z$) and energy gain by externally injected electrons. The top plot of Fig.~\ref{fig:eacc_sim} shows $W_z$ along the plasma for different $t_{seed}$. The wakefields' amplitude is evaluated on-axis, at the location of the electron bunch and is taken as the maximum value over one plasma period. The plot shows that $W_z$ is maximum for \unit[$t_{seed} = +100$ to $+200$]{ps} (and follows the same trend as the measurement in Fig.~\ref{fig:eacc}) ~\cite{EAAC2015}. As expected, this corresponds to the largest combination of seed wakefields (proportional to the beam density at $t_{seed}$) and charge driving the wakefields, $N_p$. In all cases, the maximum amplitude is reached around \unit[5]{m} into the plasma. After this point, all curve values decrease by as much as \unit[50]{\%} when reaching \unit[10]{m}~\cite{noteDENSTEP}.

Our simulation results (as well as previous work on simulations and theory \cite{SMI2,GROWTH2}) show that the phase velocity of the wakefields is changing during SM development. It is coupled to the growth rate of the SM process and is slower than the velocity of the protons (during growth) \cite{SMI2,GROWTH2} and too slow to effectively accelerate already relativistic electrons. 
The phase evolves strongly over the first \unit[$\sim5$]{m} of plasma. Because of the combined effect of (longitudinal) dephasing with respect to the wakefields and of transverse defocusing wakefields, witness electrons can only gain energy effectively in the second half of the plasma (\unit[$z>5$]{m}), after the peak of the wakefields, where their phase velocity is much more constant and close to that of the protons' velocity. This effect and the presence of a density ramp at the plasma entrance in the experiment motivated external injection at an angle with respect to the plasma column~\cite{ROYAL,AAC2018, EAAC2015, sideinj}. Simulation results show that with this injection geometry, test electrons that are eventually accelerated to the highest energy may be confined to a region outside the peak wakefields over the first few meters of plasma, and eventually drop into the wakefields when full SM of the proton bunch has occurred \cite{Wdecay}. 

The bottom plot of Fig.~\ref{fig:eacc_sim} shows the integral value of $W_z$ from \unit[5 to 10]{m} along the plasma, for each $t_{seed}$. Note that while the integral value of the wakefields depends on the integration range, the observed dependency does not. We interpret this integral (colored points on the bottom plot of Fig.~\ref{fig:eacc_sim}) as the maximum possible energy gain for electrons. 

It is clear from these results that the charge capture and acceleration processes are very intricate when injecting at an angle and in evolving wakefields. The experimental results presented here show that, for a given experimental situation, general trends are maintained, even in the presence of all these effects. These results therefore provide information about the global evolution of the wakefields in the experiment rather than about the capture and acceleration processes. We also note that injection is inherently a 3D process in the experiment, whereas in the 2D simulations all electrons always converge exactly towards the axis. Charge capture comparisons are therefore not performed here.

In simulations test electrons are injected in the wakefields over the first \unit[3]{m} of the plasma, with electrons crossing the wakefields' axis with the same shallow angle as in the experiment (\unit[$(1.5\pm0.8)$]{mrad}). Results show that some of the electrons are transported by the wakefields over the first \unit[5]{m} of plasma, but do not gain significant amounts of energy. They only gain energy from the \unit[5]{m} point on, confirming the phase velocity argument above. The electrons' maximum final energy is indicated by the gray dots on the bottom plot of Fig.~\ref{fig:eacc_sim}. The simulated electron energy gains (see bottom plot of Fig.~\ref{fig:eacc_sim}) 
follow the same trend as the integrated field values and confirm our understanding of the longitudinal wakefields driven by a self-modulating proton bunch. However, the energy gain values are about half the values obtained from the field integrals. Since we do not have access to the length over which electrons are accelerated in the experiment, we use the comparison in trend and not in value between energy gain in simulations and experiments to learn about the longitudinal wakefields. 

The electron energy values experimentally observed are between simulation values obtained from longitudinal field integrals 
and from electron energy gain. This is not surprising as electrons in general do not remain at the peak of the accelerating field within a plasma period, all along the plasma length. However, all three curves in Figs.~\ref{fig:eacc} and  \ref{fig:eacc_sim} follow the same trend as a function of $t_{seed}$. 
 
The longitudinal wakefields exhibit a behavior similar to that of the transverse wakefields (see Figs.~\ref{fig:Wr} and \ref{fig:maxr}). However, the integral value of $W_{z}$ increases up to seed points \unit[$\sim+125$]{ps}, then decreases. This is because the value of $W_{z}$ was sampled at a fixed delay by the electrons, while the maximum defocused protons can come from different positions behind the seed point (the point of the highest integrated wakefield's amplitude along the bunch).

Within the constraints of experimental measurements, trends observed in data are reasonably well described by simulations. In simulations, externally injected test electrons gain energies of several hundred MeV (see bottom plot of Fig.~\ref{fig:eacc_sim}) and follow the same trend as experimentally measured in Fig.~\ref{fig:eacc}. We observe in simulations that the final electron energy is sensitive to the electron injection angle and position, but that the measured trend as a function of $t_{seed}$ is not.

These experimental results are an important ingredient for future experiments based on the SM concept that use a self-modulation plasma section, followed by an acceleration section~\cite{bib:run2}. The results presented here are consistent with SM saturation over \unit[10]{m} of plasma, even at this low density (\unit[$n_{pe}=2\times10^{14}$]{electrons/cm$^3$}); lower density than that which gave larger energy gain (\unit[$n_{pe}=6.6\times10^{14}$]{electrons/cm$^3$}) \cite{ILOVEEA}. The results are also consistent with previous experimental studies~\cite{SSMMarlene}. 


In summary, we show that the effect of transverse wakefields from numerical simulations, integrated over the first \unit[3.5]{m} of plasma reproduces the trends observed with protons defocusing in experiment and simulations. Similarly, the effect of longitudinal wakefields on externally injected electrons and integrated over the last \unit[5]{m} of plasma reproduce the trends observed with energy gain in experiment and simulations. We observe these trends when changing the seed timing along the proton bunch and neglecting the details in exact wakefields' amplitude values or length experienced by each particle. We checked that adding these fine details to the simulation results analysis does not change this general agreement and would be beyond the claims made here.
The generally good agreement between simulations and experiment demonstrated that the development of proton bunch self-modulation in plasma is reasonably well described and understood. 
This is important for the design of accelerators based on this scheme~\cite{calwellHEP}.


\section*{Acknowledgements}
This work was supported in parts by the National Research Foundation of Korea (Nos. NRF-2016R1A5A1013277 and NRF-2019R1F1A1062377); a Leverhulme Trust Research Project Grant RPG-2017-143 and by STFC (AWAKE-UK, Cockroft Institute core and UCL consolidated grants), United Kingdom; the Russian Science Foundation, project 20-12-00062, for Novosibirsk's contribution; a Deutsche Forschungsgemeinschaft project grant PU 213-6/1 ``Three-dimensional quasi-static simulations of beam self-modulation for plasma wakefield acceleration''; the Portuguese FCT---Foundation for Science and Technology, through grants CERN/FIS-TEC/0032/2017, PTDC-FIS-PLA-2940-2014, UID/FIS/50010/2013 and SFRH/IF/01635/2015; NSERC and CNRC for TRIUMF's contribution; the U.S.\ National Science Foundation under grant PHY-1903316; the Wolfgang Gentner Programme of the German Federal Ministry of Education and Research (grant no.\ 05E15CHA); and the Research Council of Norway. M. Wing acknowledges the support of the Alexander von Humboldt Stiftung and DESY, Hamburg. Support of the Wigner Datacenter Cloud facility through the "Awakelaser" project is acknowledged.  The work of V. Hafych has been supported by the European Union's Framework Programme for Research and Innovation Horizon 2020 (2014--2020) under the Marie Sklodowska-Curie Grant Agreement No.\ 765710.  The AWAKE collaboration acknowledge the SPS team for their excellent proton delivery.


\begin{thebibliography}{99}
\bibitem{EXP1} The absence of a $\xi_{max}$ shift for \unit[$t_{seed} < 100$]{ps} is the result of the transverse wakefields being quite similar for \unit[$t_{seed} \geq 0$]{ps} but changing for \unit[$t_{seed} < 0$]{ps} (see top plot of Fig. \ref{fig:Wr}). The exact position of $\xi_{max}$ depends on the wakefields phase evolution during the growth of self-modulation. It is generally expected to shift to larger values for increasing $t_{seed}$.
\bibitem{noteDENSTEP}We note here that setting a plasma electron density step at the percent-level close to the plasma entrance was proposed for the wakefields to maintain values near their saturation value~\cite{DENSTEP}. This is essential for acceleration over long distance and large energy gain.
\end{thebibliography}
\end{document}